\documentclass[aps,pra,twocolumn,showpacs]{revtex4-1}

\usepackage{graphicx}
\usepackage[export]{adjustbox}
\usepackage{dcolumn}
\usepackage{bm}
\usepackage{color}
\usepackage{amssymb}
\usepackage{amsmath}
\usepackage{url}
\usepackage{epstopdf}
\usepackage{hyperref}
\usepackage{comment}
\usepackage[caption=false]{subfig}
\usepackage{tabularx,ragged2e}
\usepackage{booktabs}
\usepackage[percent]{overpic}
\usepackage{soul}
\usepackage{amscd}
\usepackage{bm}
\usepackage{xcolor}

\usepackage{tikz}
\usetikzlibrary{arrows}

\newcommand*\pct{\scalebox{.9}{\%}}

\newcommand{\co}[1]{{#1}} 
\setstcolor{red}    

\begin{document}


\title{Multielectron effects in strong-field ionization of benzene derivatives}

\author{Eldad Yahel}
\email[]{yahel@eng.tau.ac.il}
\author{Amir Natan}
\altaffiliation{The Sackler Center for Computational Molecular and Materials Science, Tel-Aviv University, Tel-Aviv 69978, Israel}
\email[]{Corresponding Author: amirnatan@post.tau.ac.il}

\affiliation{Department of Physical Electronics, Tel Aviv University, Tel Aviv, Israel.}


\date{\today}

\begin{abstract}
\noindent Strong-field ionization of benzene, fluorobenzene, benzonitrile and 1-chloro-2-fluorobenzene is studied within the framework of real-space and real-time time-dependent density functional theory. Analysis of the ionization rates as a function of the molecule orientation reveals a significant multielectron contribution from inner Kohn-Sham orbitals that depends on the electronic structure and on the orbital symmetries of the molecule, as well as on the polarization and intensity of the external laser field. Calculated photoelectron angular distributions at different molecular orientations and in response to laser fields with different degrees of ellipticity further demonstrate the spatial dependency of the orbital ionization rates. 
\end{abstract}


\maketitle

\section{Introduction}
\noindent The nonlinear optical response of polyatomic molecules to strong laser fields has been the subject of \co{extensive} research in recent years \cite{Hay2000a, Holmegaard2010, Hansen2011c, Hansen2011, Dimitrovski2015, Ferre2015}. The interaction between a short and strong laser field and a molecule excites the highest electrons and induces ionization in the direction of the laser polarization. This process is more efficient if the molecules are pre-aligned, and depends on the symmetry properties of the ionized electronic orbitals and on the relative molecular orientation with respect to the laser polarization \cite{Hansen2011, Hansen2011c}. Furthermore, in comparison to atoms, the extra degrees of freedoms of the molecule result in contributions from multiple correlated ionization channels \cite{Ferre2015}. Typically, the total electron ionization is measured at different molecular orientations \cite{Pavicic2007}. In addition, the photoelectron angular distributions (PAD) provide information on the emission direction of the electrons. \co{The PAD spatial dependence further demonstrates the symmetry properties of the electronic orbitals and their relative contributions to the total molecular ionization \cite{Hansen2011, Holmegaard2010}. For non-linearly polarized laser fields, the interaction of the ionized electrons with the nucleus is significantly minimized. This facilitates the reconstruction of the outer-most molecular orbital structure \cite{Itatani2004}.}\\
\indent Benzene is a planar symmetric-top structure, which is characterized by the $\mathrm{D}_{6h}$ point symmetry group \cite{Demaison2013b}. In substituted benzene molecules, one or more of the hydrogen atoms are replaced by another ion or functional group, e.g. fluoride, chlorine or carbon-nitrogen, resulting in an asymmetric-top structure with point-group symmetry that is either $\mathrm{C}_{2\nu}$ (\co{fluorobenzene}, benzonitrile) or reduced to $\mathrm{C}_s$ only (1-chloro-2-\co{fluorobenzene}) \cite{Hansen2011, Demaison2013b, Kjeldsen2005, Wohlfart2008}. This structural change results in a different electronic configuration, deformation of the benzene ring and redistribution of the electronic charge within the occupied orbitals \cite{Kjeldsen2005, Demaison2013b}. This alters the symmetry properties of the molecular orbitals, which can have a profound effect on their ionization dynamics \cite{Son2009f}. The influence of the electronic structure on the ionization properties of substituted benzene was mostly studied based on simplifications of the time-dependent Schr\"{o}dinger equation (TDSE) or on approximated analytical methods, in order to calculate the orientation-dependent ionization rates and PADs \cite{Kjeldsen2005, Martiny2010, Holmegaard2010, Hansen2011, Hansen2011c}. These calculations demonstrate that in order to properly determine the ionization properties of substituted benzene, it is important to take into account the Stark shift due to the interaction of the laser field with the permanent and induced dipole moment contributions \cite{Holmegaard2010, Hansen2011}. We note, however, that in these studies, either only the \co{highest-occupied molecular orbital} (HOMO) is assumed to contribute to the ionization channels, or the contributions from multiple orbitals is added incoherently, i.e. neglecting all-important electron correlation effects \cite{Ferre2015}. On the other hand, multielectron effects are expected to be significant for benzene derivatives, due to the relatively high-level density of the molecular orbitals \cite{Otobe2007}. Computational methods based on real-time, time-dependent density-functional theory (RT-TDDFT) make it possible to properly account for the exchange and correlation of inner-shell valence electrons and their contribution to the ionization process \cite{Telnov2009h}. \co{This approach was employed to study the ionization rates in benzene and the orientation-dependent contribution of degenerate HOMOs \cite{Otobe2007, Dundas2012, Wardlow2015b}.} \\
\indent Our goal in this work is to further examine the dependency of the ionization rates on the relative orientation between the \co{molecular axes} and the laser polarization for different laser intensities. \co{We also aim to study} the effect of laser polarization and ellipticity on the angular ionization pattern, in both symmetric and asymmetric benzene derivatives. In particular, we wish to analyze the contributions of inner-shell valence molecular orbitals to the total \co{molecular ionization}, and the effect of the breakdown of orbital symmetry. We base our calculations on the Bayreuth version of the PARSEC code \cite{Mundt2007c,Mundt2009d}, which is efficiently parallelized using sub-domain decomposition methods \cite{Chelikowsky1994c,Kronik2006a}. \co{This approach facilitates an efficient representation of} both localized and delocalized continuum states on a real-space grid, and is well suited for strong-field physics \cite{Mundt2009d}. \co{We limit our analysis to femto-second laser pulses with a few optical cycles and to the Born-Oppenheimer approximation. In addition, we assume that the atoms are confined to their positions.} Our results demonstrate strong ionization asymmetries, depending on the external laser ellipticity and \co{on} the dipole of the molecule, as well as significant contributions from inner-shell valence orbitals, at certain laser polarizations and strong-enough intensities.\\
\indent This work is organized as follows. In Sec. II, we review the theoretical framework for the TDDFT calculations and state our basic assumptions. In Sec. III, we present the ground-state configurations of selected molecular orbitals that are \co{important} for the understanding of the ionization dynamics, and discuss a few examples of numerical calculations of the ionization rates and photoelectron angular distributions. To this end, we consider benzene, fluorobenzene, benzonitrile and 1-chloro-2-fluorobenzene molecules, in which the substituted benzene functional group results in somewhat different electronic structure, orbital symmetry, permanent electronic dipole and polarization properties. In particular, we study the effects of the fixed-in-space molecular orientation, laser polarization, ellipticity and intensity on the calculated orbital and total ionization yields, as well as on the angular ionization distribution pattern. Finally, in Sec IV, we provide some concluding remarks.
\section{computational method}
\noindent 
We model the electronic response of a molecule to strong-field laser interaction within the framework of real-space and real-time propagation TDDFT. Here, we shortly review the formalism for completeness. \co{It is assumed that only valence electrons contribute significantly to the excited molecule charge density, and that the external field has a short time duration and weak enough intensity to justify neglecting molecular dynamics and double ionization effects \cite{Hay2000}.} Employing the ground state Hamiltonian, $\hat{H}_0$, the Kohn-Sham (KS) equation for the $k$-th state orbital, $\varphi_k(\vec{r})$, is given as, 
\begin{align}
\hat{H}_{0}[\rho,\mathbf{R}_a] \varphi_k &\equiv \Big[ -\frac{1}{2} \nabla^2 + \nu_H[\rho(\mathbf{r})] \nonumber \\ 
+ \sum_a \hat{\nu}_{ps,a}[ (\mathbf{r}-\mathbf{R_a}) ] &+ \nu_{xc}[\rho(\mathbf{r})] \Big] \varphi_k = \epsilon_k \varphi_k. \label{H_0}
\end{align}
\noindent Here, atomic units are used throughout and $\rho(\mathbf{r})=2\sum _{k=1}^{N_e}|\varphi _{k}(\mathbf{r})|^2$ is the electronic density of $N_e$ non-spin polarized electronic states with energy $\epsilon_k$. \co{Other terms in (\ref{H_0}) are the Hartree potential, $\nu_H[\rho(\mathbf{r})] = \int\rho(\mathbf{r'})/|\mathbf{r}-\mathbf{r'}|\mathrm{d}\mathbf{r'}$, and the exchange-correlation potential approximation, $\nu_{xc}$ \cite{VanLeeuwen1998}}. The ionic pseudopotential $\hat{\nu}_{ps,a}$ for an atom $a$ at fixed position $\mathbf{R_a}$ is calculated using norm conserving pseudopotentials and the Kleinman-Bylander projection for the non-local part \cite{Troullier1991d, Kleinman1982b}. In the real-space representation, the orbitals are sampled on a uniform grid and the differential operator in (\ref{H_0}) is replaced by its high-order finite difference equivalent form \cite{Chelikowsky1994c}. In the time-dependent calculations, one starts from the ground-state electronic density and solves the multielectron response to an external electro-magnetic field according to the system Hamiltonian, $\hat{H}[\rho,\mathbf{r},t] = \hat{H}_0[\rho(\mathbf{r},t),\mathbf{R}_a] + \hat{H}_{ext}$, where $\hat{H}_{ext}$ is the external-field Hamiltonian. That is, the density $\rho(\mathbf{r})$ is replaced with the time-dependent density $\rho(\mathbf{r},t)$, and this immediately affects the terms $\nu_H$ and $\nu_{xc}$, where for the latter we use the adiabatic approximation \cite{Mundt2009d}. The propagated orbitals satisfy the time-dependent KS equations \cite{Runge1984},
\begin{equation}
i\frac{\partial}{\partial t} \varphi_{k}(\mathbf{r},t) = \hat{H}[\rho(\mathbf{r},t),\mathbf{r},t] \varphi_{k}(\mathbf{r},t).
\label{KS}
\end{equation}
The real-time TDDFT solution of (\ref{KS}) in the strong-field regime is carried out without linearization, and it is composed of two steps. First, the ground-state KS orbitals of the molecule are calculated self-consistently within the framework of time-independent DFT, without applying an external field. The static solutions are then used as initial conditions for the subsequent time-dependent solution, i.e. $\varphi_{k}(\mathbf{r},t = 0)$. In this step, we switch on the laser adiabatically and propagate the KS orbitals explicitly in real time, under strong-field conditions. For the propagation scheme we employ a $4^{th}$-order Taylor expansion for approximating the propagator $\hat{U}(t, t+ \Delta t)$, between time $t$ and $t+\Delta t$, i.e. \cite{Mundt2009d}:
\begin{align}
&\varphi_{k}(\mathbf{r},t+\Delta t)=\hat{U}(t,t+\Delta t)\varphi_{k}(\mathbf{r},t) \notag \\
&\simeq \sum\limits_{n=1}^{4} \frac{1}{n!}\left[-i\Delta t \hat{H}(\mathbf{r},t+\Delta t/2)\right]^n \varphi_{k}(\mathbf{r},t).\label{propagator}
\end{align}
\indent In this explicit approach, one assumes that the Hamiltonian is approximately time-independent for a short enough integration time of $\Delta t$, and it is calculated at the mid-point of the integration interval, i.e. $\hat{H}(\mathbf{r},t+\Delta t/2) \simeq 1/2 \times [\hat{H}(\mathbf{r},t) + \hat{H}(\mathbf{r},t+\Delta t)]$ \cite{Mundt2009d}. The orbitals are integrated in time using a predictor-corrector scheme \co{that} is typically satisfied with a single iteration. While this method is not strictly mathematically accurate, it was shown to conserve the orbital norms to a good level of approximation \cite{Castro2004b}. In this scheme, all the occupied valence orbitals are propagated in time. \co{This} essentially provides the multielectron response of the molecule to the external laser field, including the screening effect beyond the single-active electron approximation. Furthermore, the KS equations automatically account for the orientation-dependent Stark shift effect on the propagating orbitals \co{that must be taken into account for attaining an accurate description} of the ionization process in molecules with large static dipole moments and polarizabilities \citep{Abu-samha2010b, Kornev2016}. We restrict our calculations to a spherical domain \co{that} is big enough to accommodate the time-dependent electronic orbitals under the effect of the external field, and enables convergence of the static solutions in the initial step. Furthermore, in order to mimic outgoing boundary conditions and prevent non-physical reflections of the ionized electronic density from the domain boundaries, we employ an absorbing layer near the surface of the computational edge. \co{This condition} is realized by multiplying the propagating orbitals within this layer with a mask function \cite{Reinhard2006}. \co{Furthermore, the} adiabatic elimination of orbital occupation at the boundary layer provides a measure for the unbound states that propagate away from the molecule during the ionization process. Accordingly, we define the total ionization probability as $P(t) \equiv 1 - \mathop{\Pi}_{k}\left[1-P_{k}(t)\right]$, where $P_{k}(t)=1-N_{k}(t)$ is the $k$'s orbital ionization probability and $N_{k}(t) = \int \varphi_{k}^*(\mathbf{r},t) \varphi_{k}(\mathbf{r},t) \mathrm{d}^3r$ accounts for the residual orbital normalization that decreases in time due to the absorbing boundary conditions \cite{Son2009f}.\\
\indent The choice of exchange-correlation potential, $\nu_{xc}$, can dramatically affect the results. Specifically, the local density approximation (LDA) might not be accurate due to the self-interaction error that also results in a wrong asymptotic tail behaviour of the potential as $r \rightarrow \infty$. This effect results in lower ionization potential for the HOMO, and, therefore, in over-estimated ionization rates \cite{Otobe2007}. Thus, care should be taken in the choice of functional in order that the results will be commensurate with experimental measurements, in particular for the vertical ionization potential \cite{Chong2002}. \co{In this work we employ a modified form of the long-range asymptotically-corrected Leeuwen-Baerends (LB) exchange-correlation functional, which properly accounts for the tail of the effective potential away from the ions \cite{Schipper2000a}}. This approach was recently found to provide good results in predicting the ground-state energies and ionization response of the HOMO, HOMO$-1$ and HOMO$-2$ orbitals in simulations of diatomic and linear molecules \cite{Son2009f, Telnov2009h}. To this end, the system-dependent parameters $\alpha$ and $\beta$ \co{in the modified LB approximation} are selected so that the absolute ground-state HOMO energy is in good agreement with the measured ionization potential \cite{Telnov2009h}. In the case of benzene derivatives, we found the functional parameters $\alpha = 1.0$ and $\beta = 0.01$ to be optimal. \\
\indent We further assume the dipole approximation, as the dimensions of the molecule system are very small \co{compared to} the external field wavelength, and, hence, the spatial dependency of the field can be neglected. Under this condition, the perturbative part of the Hamiltonian that accounts for the external time-dependent laser field is given by $\hat{H}_{ext}=\mathbf{\mathcal{E}}(t) \cdot \mathbf{r}$ in the length gauge \cite{Dundas2012, Telnov2009h}. Here, the external time-dependent laser electric field $\mathbf{\mathcal{E}}(t)$ is
\begin{align}
\mathbf{\mathcal{E}}(t) &= \mathcal{E}_0(t)\Big[ \frac{\epsilon}{\sqrt{1+\epsilon^2}}\color{blue} \mathbf{\hat{e}_i} \color{black} \sin (w_0 t + \nu) \nonumber \\
&+ \frac{1}{\sqrt{1+\epsilon^2}}\color{blue}\mathbf{\hat{e}_j} \color{black} \cos (w_0 t + \nu) \Big],\label{external} 
\end{align}
where $\epsilon$ is an ellipticity parameter, $w_0$ is the carrier frequency and $\nu$ is the carrier envelope phase. \co{Thus} $\epsilon = 0$ for a linearly polarized field, $\epsilon = \pm 1$ for left (+) and right (-) circularly polarized fields, and intermediate values of $\epsilon$ represent general elliptic polarizations. \co{The unit vectors $\mathbf{\hat{e}_i}$ and $\mathbf{\hat{e}_j}$} correspond to orthogonal polarization directions in coordinate space. We assume a sine-squared pulse shape, so that the pulse envelope is given by
\begin{equation}
\mathcal{E}_0(t) = \mathcal{E}_0 \sin^2\left( \frac{\pi t}{N T} \right),
\end{equation}
where $T$ is the pulse duration, $N$ is the number of optical cycles and $\mathcal{E}_0$ is the peak field strength.
\begin{figure}[b]
\centering
\captionsetup[subfigure]{labelformat=empty,position=top,labelfont=bf,textfont=normalfont,singlelinecheck=off,justification=raggedright}
\renewcommand{\thesubfigure}{A}
\subfloat[]{\label{first}%
   \includegraphics[width=0.80\linewidth]{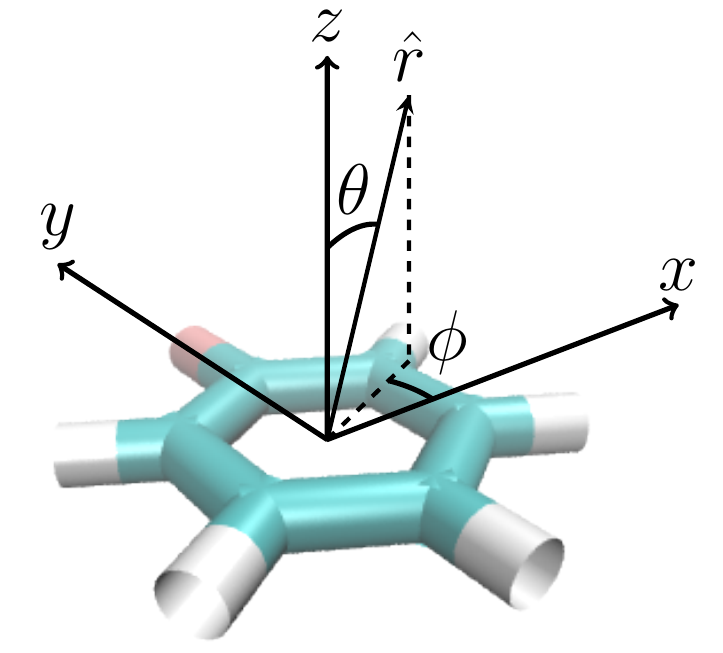}%
}
\caption{Molecule and photoemission geometry that is assumed in this work. \co{The principle axes} are aligned along the fixed-body frame ($x$,$y$,$z$), where the direction $y$ is along the most polarizable \co{molecular axis} and points towards the functional group. The angles $\theta$ and $\phi$ denote the spherical coordinates \co{towards the direction $\mathbf{\hat{r}}$ in space.}\\}\label{coordinates}
\end{figure}

\indent We consider molecules with a planar-top geometry \co{that} are either rotationally symmetric (benzene) or asymmetric (fluorobenzene, benzonitrile and 1-chloro-2-fluorobenzene). We use a three-dimensional Cartesian grid and define the molecular (body-fixed) frame ($x$,$y$,$z$) so that the \co{$x$ and $y$ axes} originate from the center-of mass and lie in the \co{molecular plane}, as illustrated in Fig. \ref{coordinates}. \co{Our convention is that the most-polarizable principle axis of the molecule is along the $y$ direction, and that the head and tail of the molecule correspond to the positive and negative directions of that axis, respectively.} For a fixed laser polarization, we compute the PADs from the TDDFT solutions, which makes it possible for us to visualize the angular differential ionization probability of electrons in the \co{molecular frame} \cite{Kjeldsen2006b, Son2009f}. The orbital-dependent angular differential ionization probability through a solid angle $\mathrm{d} \Omega = \sin\theta\,d\theta \,d\phi$ in \co{direction $\mathbf{\hat{r}}(\phi, \theta)$} is defined in terms of the flux flowing through the surface of a sphere of radius $r_0$ \cite{Son2009f}:
\begin{equation}
\frac{\partial P_{k}(t)}{\partial \Omega} = \int_0^t r_0^2 \cdot \mathrm{Im} \left [ \varphi_{k}^*(\mathbf{r},\tau) \frac{\partial}{\partial r}  \varphi_{k}(\mathbf{r},\tau) \right ]_{r=r_0} \mathrm{d}\tau,\label{PAD}
\end{equation}
\noindent and the total angular differential ionization probability through this sphere is given by the contribution of all orbitals, i.e. $\partial P(t) / \partial \Omega = 1- \mathop{\Pi}_{k}(1-\partial P_{k}(t) / \partial \Omega)$. Here, the polar angle $\theta$ is defined between the \co{molecular $z$ axis} and the direction $\mathbf{\hat{r}}$, whereas $\phi$ is the azimuth angle between the \co{molecular $x$ axis} and the projection of $\mathbf{\hat{r}}$ in the \co{molecular plane.}\\
\newcommand{\mysize}{0.15}
\nocite{Otobe2007}
\nocite{gilbert1971}
\nocite{Hansen2011c}
\nocite{mohraz1980}
\begin{figure*}
\captionsetup[subfigure]{labelformat=empty,position=top,labelfont=bf,textfont=normalfont,singlelinecheck=off,justification=raggedright}
\centering
\subfloat[]{\label{second}%
  \includegraphics[width=\linewidth]{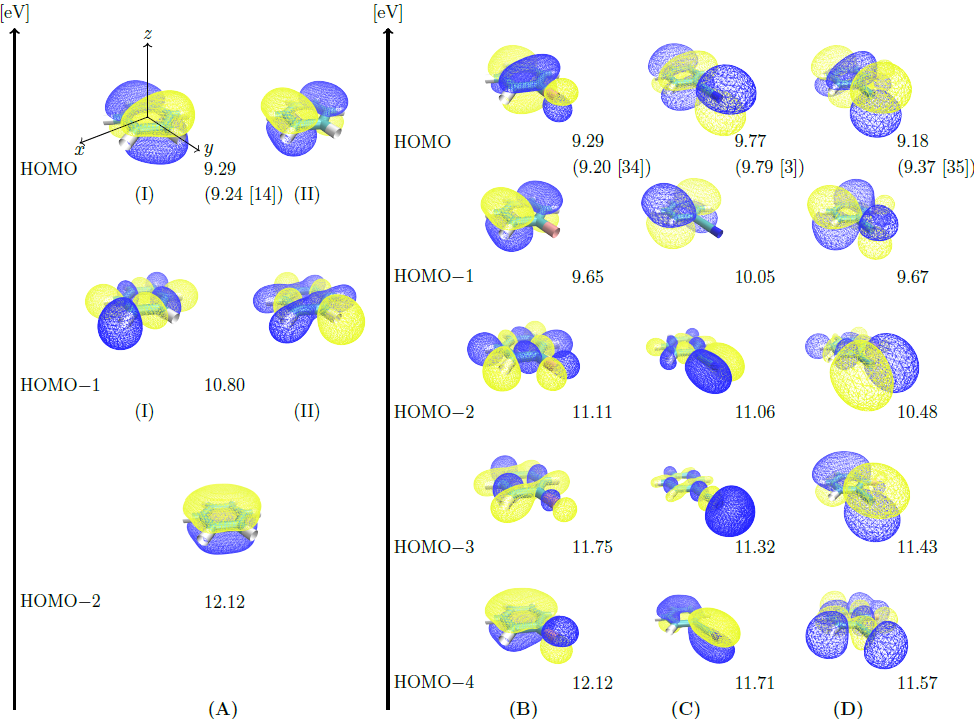}%
}
\caption{Molecular bonds and isocontour plots of selected Kohn-Sham molecular orbital distributions \co{for (A) benzene, (B) fluorobenzene, (C) benzonitrile and (D) 1-chloro-2-fluorobenzene.} Pseudo-colors correspond to different signs of the orbital function, whereas the absolute energies of the respective Kohn-Sham orbitals (and the measured first ionization energies) are given in units of electro-volts (eV).\\}\label{orbitals}
\end{figure*}

\section{results and discussion}
\subsection{Ground-state Configurations}
\noindent For each molecule, we performed a geometrical structure relaxation with LDA on a fine grid ($\Delta x = \Delta y = \Delta z \sim 0.2$ a.u.), whereas all other calculations were done using the LB functional on a coarser grid ($\sim 0.4$ a.u.) that still provided satisfactory convergence of the orbital solutions. \co{This} choice of the functional approximation results in different absolute ionization yields compared to LDA; however, we found that it does not change the ground-state orbital ordering, and that a qualitatively similar response is achieved using \co{both approximations} for different functional group substitutions and molecular orientations. In most simulations we used a spherical domain with $R \sim 40$ a.u. \co{that} was large enough to converge the ionization rates within $\approx 1 \pct$ accuracy. \co{A temporal resolution of $0.001$ femto-seconds was found to \co{ensure} the stability of the propagated orbitals.}\\
\indent \co{In Fig. \ref{orbitals} we examine the symmetry properties of the highest-occupied molecular orbitals of the benzene derivatives that are analyzed in this work. The highest-occupied molecular orbitals are anti-symmetric upon reflection in the molecular plane. For the symmetric-top benzene molecule, these orbitals are also either symmetric or anti-symmetric with respect to reflection in the $xz$ and $yz$ planes. Here, the HOMO is a two-fold degenerate $\pi$ orbital, whereas the HOMO$-1$ is a doubly-degenerate $\sigma$ orbital \cite{Otobe2007}. The HOMO$-2$ is a $\pi$ orbital with zero nodes, besides the one that lies in the molecular plane. Hence, the electronic charge distribution of benzene is symmetric with a \co{zero permanent dipole moment}. In the case of asymmetric top molecules, the substitution of the functional group breaks the orbital symmetries, which lifts the orbital degeneracy and leads to a non-zero permanent dipole moment \citep{Abu-samha2010b}; Table \ref{parameters} summarizes the calculated static dipole moments, i.e. $\bm{\mu}$, of asymmetric-top benzene substitutes, along with their corresponding experimental values. The HOMO is characterized by a nodal surface that is approximately parallel to the $xz$ plane, but is no longer anti-symmetric with respect to reflection in the $xz$ plane, as part of the orbital density is localized on the substituted atom. In particular, in fluorobenzene the asymmetry in the electronic charge distribution along the principle $y$ axis is relatively small, whereas in benzonitrile it is much enhanced, resulting in a larger net permanent dipole for this molecule. Furthermore, the HOMO of fluorobenzene is characterized by a similar ionization potential to that of benzene, whereas in benzonitrile the electronic reconfiguration results in a more strongly bound HOMO compared to the other benzene derivatives that we discuss in this work. The HOMO$-1$ of fluorobenzene and benzonitrile is anti-symmetric upon reflection in its $yz$ nodal plane and symmetric upon reflection in the $xz$ plane, whereas the next-order occupied orbitals are characterized by an asymmetrical distribution with respect to reflection in the $xz$ plane. On the other hand, the case of 1-chloro-2-fluorobenzene is somewhat different, as both the HOMO and HOMO$-1$ orbitals are strictly anti-symmetric only with respect to the molecular plane. Nonetheless, these orbitals have nodal planes that are approximately parallel to the $xz$ and $yz$ planes, respectively \cite{Kjeldsen2005}. Here, the molecular orbital density is partly localized at the sites of the substituted chlorine and fluoride atoms, and, hence, the direction of the net dipole is not effectively along one of the fixed-body axes.}
\begin{table}[h]
  \caption{Calculated and measured magnitudes of non-negligible dipole components for the asymmetric-top benzene substitutes, in units of Debye. Note that for 1-chloro-2-fluorobenzene ($\mu_x$, $\mu_y$) is shown.}
\Centering
  \begin{tabularx}{\linewidth}{c *2{>{\Centering}X}}\toprule
    & $\bm{\mu}$ [calculated] & $\bm{\mu}$ [measured] \\
    \hline\hline
    fluorobenzene & 1.76 & 1.66 \cite{Kurtz1990}  \\
    benzonitrile & 4.68 & 4.52 \cite{Wohlfart2008}  \\
    1-chloro-2-fluorobenzene & ($2.17,1.15$) & ($2.28, 1.47$) \cite{Onda1994}  \\
    \hline\hline
  \end{tabularx}
  \label{parameters}
\end{table}
\subsection{Strong-field Ionization}
\begin{figure}[b]
\captionsetup[subfigure]{labelformat=empty,position=top,labelfont=bf,textfont=normalfont,singlelinecheck=off,justification=raggedright}
\subfloat[]{\label{first}%
  \includegraphics[width=\linewidth]{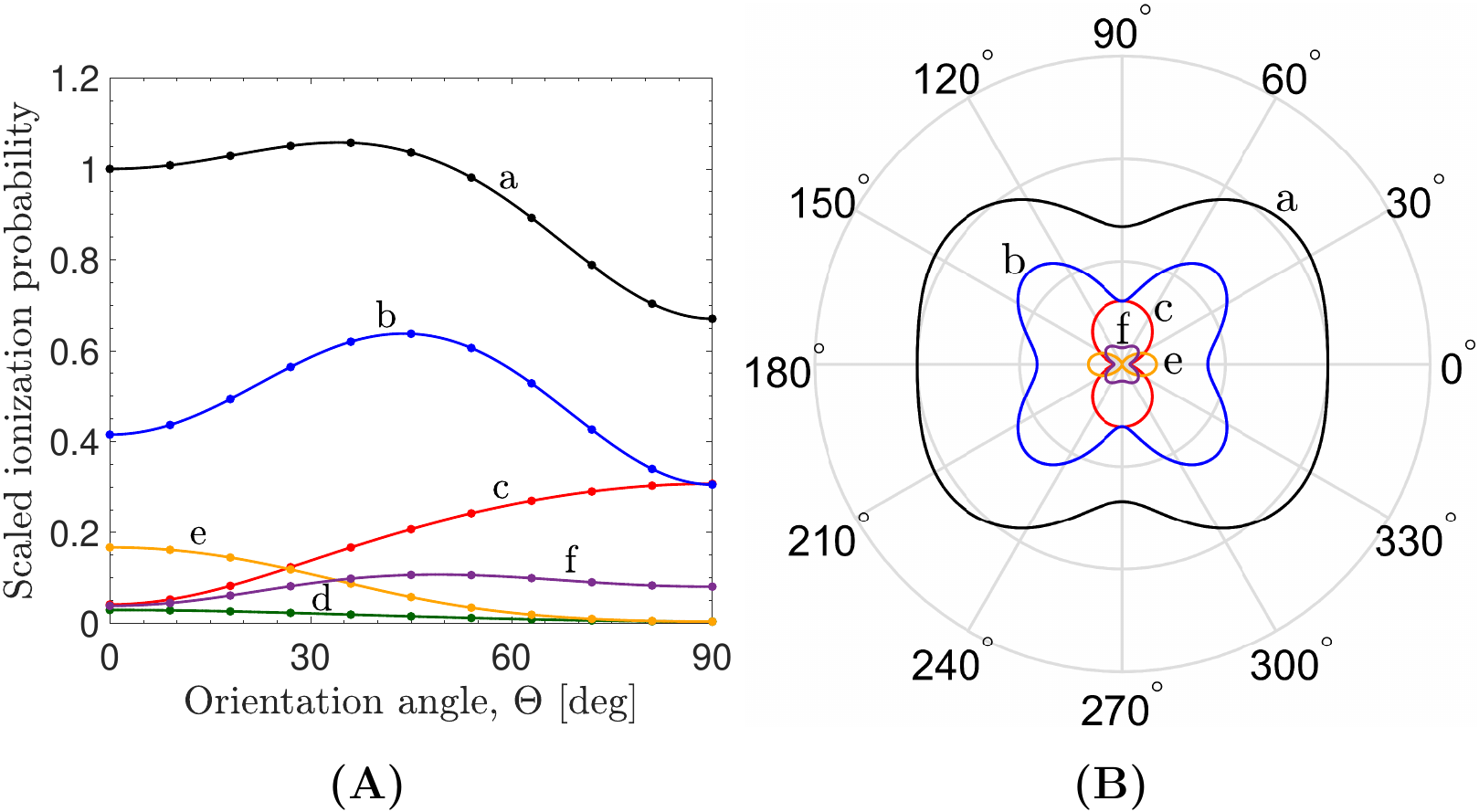}%
}
\caption{\co{Scaled ionization probabilities of benzene molecule, as a function of the orientation angle $\Theta$, for a linearly polarized laser in the $yz$ plane, with peak intensity of $1 \times 10^{14}$ W/cm$^2$, in Cartesian (panel A) and Polar (panel B) coordinates: (a) total, (b) HOMO(I), (c) HOMO(II), (d) HOMO$-1$(I), (e) HOMO$-1$(II) and (f) HOMO$-2$ ionization probabilities.}\\}\label{benzene_ionization}
\end{figure}
\begin{figure}
\centering
\captionsetup[subfigure]{labelformat=empty,position=top,labelfont=bf,textfont=normalfont,singlelinecheck=off,justification=raggedright}
\subfloat[]{\label{first}%
  \includegraphics[width=\linewidth]{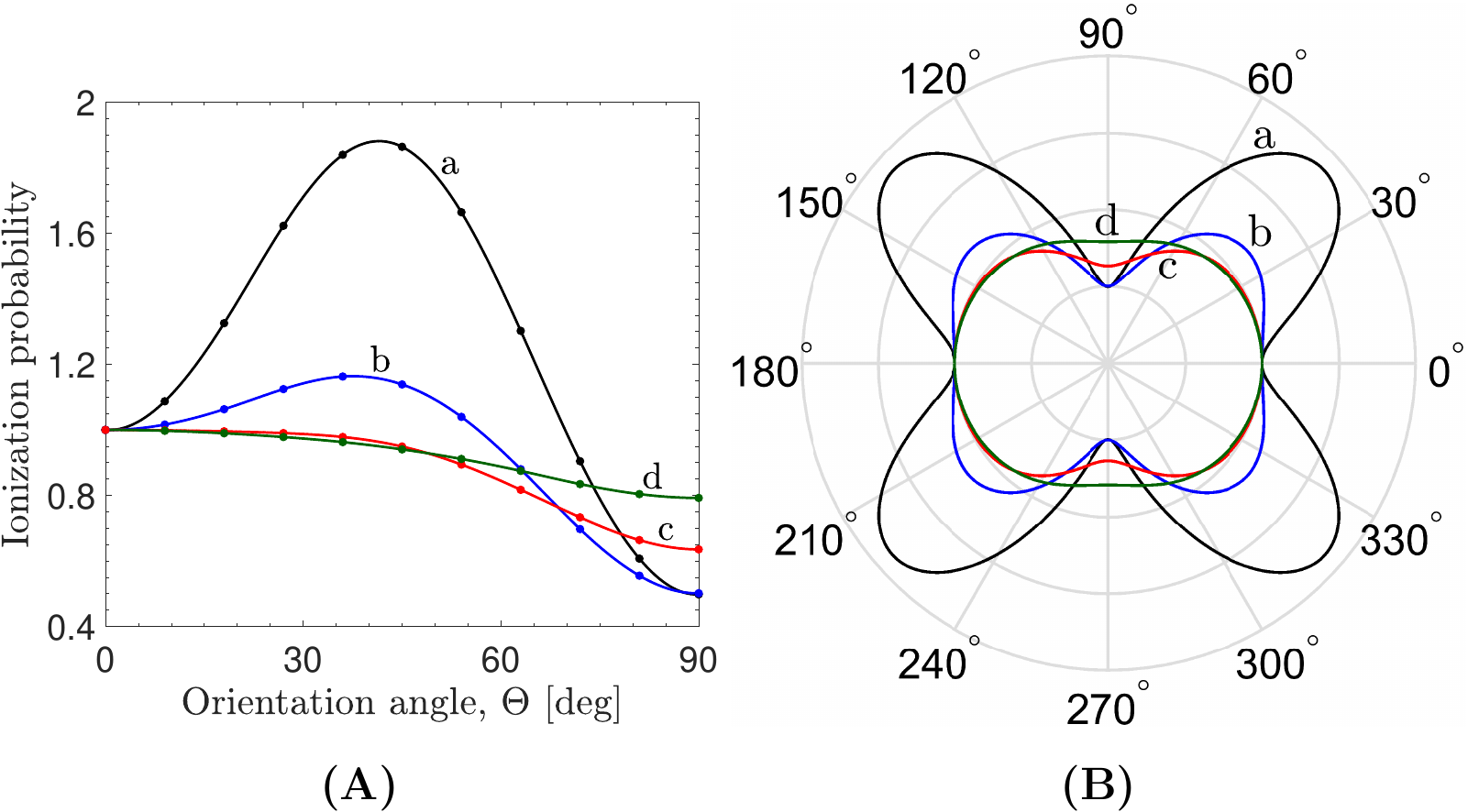}%
}
\caption{Ionization probabilities of fluorobenzene molecule, as a function of the \co{orientation angle $\Theta$, for a linearly polarized laser in the $yz$ plane with different intensities,} in Cartesian (panel A) and Polar (panel B) coordinates: (a) $2.5 \times 10^{13}$ W/cm$^2$, (b) $5 \times 10^{13}$ W/cm$^2$, (c) $1 \times 10^{14}$ W/cm$^2$ and (d) $2 \times 10^{14}$ W/cm$^2$. Here, \co{the data is scaled to the respective ionization rate at $\Theta = 0^\circ$.}}\label{fluorobenzene_ionization_intensities}
\end{figure}
\begin{figure}
\centering
\captionsetup[subfigure]{labelformat=empty,position=top,labelfont=bf,textfont=normalfont,singlelinecheck=off,justification=raggedright}
\subfloat[]{\label{first}%
  \includegraphics[width=\linewidth]{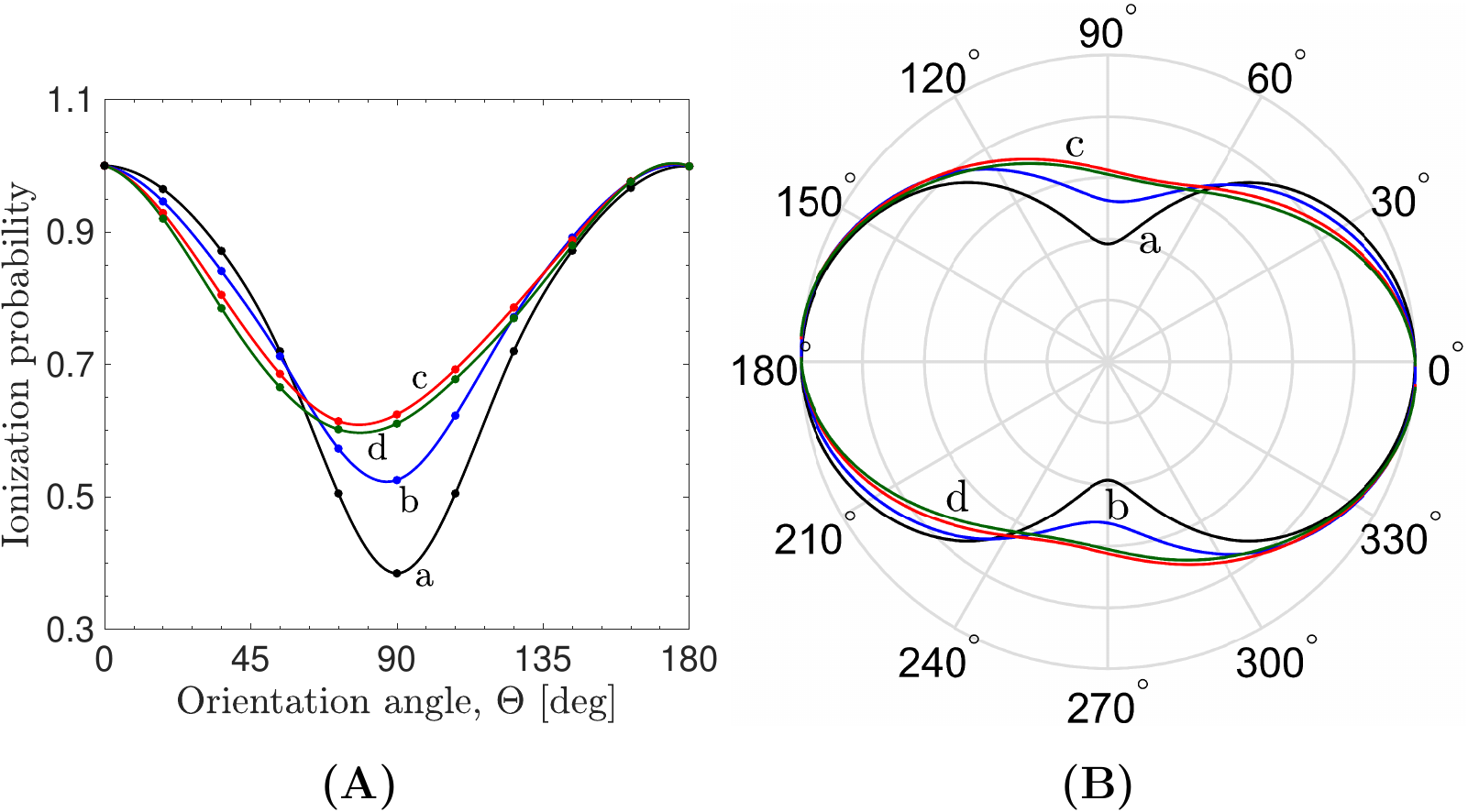}%
}
\caption{\co{Ionization probability of 1-chloro-2-fluorobenzene molecule, for laser peak intensity of $1 \times 10^{14}$ W/cm$^2$, in Cartesian (panel A) and Polar (panel B) coordinates, as a function of the orientation angle $\Theta$ between the linear laser polarization and the $y$-axis. Here, we consider different polarization planes that are defined by the directions (a) $\mathbf{\hat{r}}(\phi=0^\circ,\theta=90^\circ)$, (b) $\mathbf{\hat{r}}(\phi=0^\circ,\theta=30^\circ)$, (c) $\mathbf{\hat{r}}(\phi=0^\circ,\theta=60^\circ)$ and (d) $\mathbf{\hat{r}}(\phi=0^\circ,\theta=90^\circ)$ and the fixed $y$-axis. The data is scaled to the respective ionization rate at $\Theta = 0^\circ$.}}
\label{chlorofluorobenzene_ionization}
\end{figure}

\noindent \co{Figure \ref{benzene_ionization} depicts the scaled ionization probabilities of a fixed-in-space benzene molecule, as a function of an orientation angle, which is conveniently defined as $\Theta \equiv 90^\circ-\theta$. We consider laser polarizations in the $yz$ plane, i.e. in the direction $\mathbf{\hat{r}}(\phi=90^\circ, \Theta)$.} The laser parameters are of a sine-squared pulse shape with a wavelength of $\lambda = 800$ nm, peak intensity of $1 \times 10^{14}$ W/cm$^2$ and duration of 10 optical cycles (i.e. $\sim 26.7$ fsec). \co{For convenience, we scale the ionization rates by the total ionization rate when the laser is linearly polarized parallel to the molecular plane, i.e. in the direction $\Theta = 0^\circ$. We assume that the carrier envelope phase effects are small, and, hence, we may exploit the axial symmetry of the molecule in order to limit the ionization calculations to the range of $\Theta$ between $0^\circ$ and $90^\circ$.} Here, we consider eleven distinct orientations or a step size of $9$ degrees (scattered circular points), whereas the solid lines are spline interpolated. In principle, the optical response depends on the ionization potential of the electronic orbitals (which is taken as the negative of the KS eigenvalues), and therefore the HOMO ionization is the strongest. The interaction of the applied electric field with the electronic charge distribution within each orbital is symmetry dependent, and the angle-dependent ionization rate reflects this property \cite{Zhao2011b}. \co{We note that the HOMO(II) orbital response is effectively suppressed when the laser polarization is parallel to the molecular plane, whereas this orbital response is equal to the HOMO(I) orbital when the laser polarization is perpendicular to the molecular plane ($\Theta = 90^\circ$), in agreement with previous TDDFT calculations \cite{Dundas2012}.} That is, in the perpendicular configuration, the laser polarization is parallel to the nodal planes of both degenerate HOMOs, cf. Fig. \ref{orbitals}. The contribution from inner $\sigma$ \mbox{HOMO$-1$} orbitals is maximized when the laser polarization direction is along the \co{molecular plane}, in agreement with static DFT calculations \cite{Otobe2007}. However, for the laser intensity considered herein, the contribution of these inner orbitals is not significant, and the total ionization probability closely resembles the response of the HOMO, in particular at the larger orientation angles. The pattern of the orientation dependence resembles a center-fat propeller shape with a maximum response that is obtained at an angle of \co{$\Theta \simeq 40^\circ$}, which is quite similar to the ionization response of other molecules with $\pi$ HOMOs, e.g. CO$_2$ \cite{Son2009f}. The total ionization yield is minimized when the laser polarization is perpendicular to the \co{molecular plane} \cite{Otobe2007}. We note that an essentially similar ionization dependency is obtained if the laser is linearly polarized along the $xz$ plane, as the degenerate orbitals of benzene are symmetric \co{with respect} to an exchange of these axes.\\
\begin{figure}
\centering
\captionsetup[subfigure]{labelformat=empty,position=top,labelfont=bf,textfont=normalfont,singlelinecheck=off,justification=raggedright}
\subfloat[]{\label{first}%
  \includegraphics[width=\linewidth]{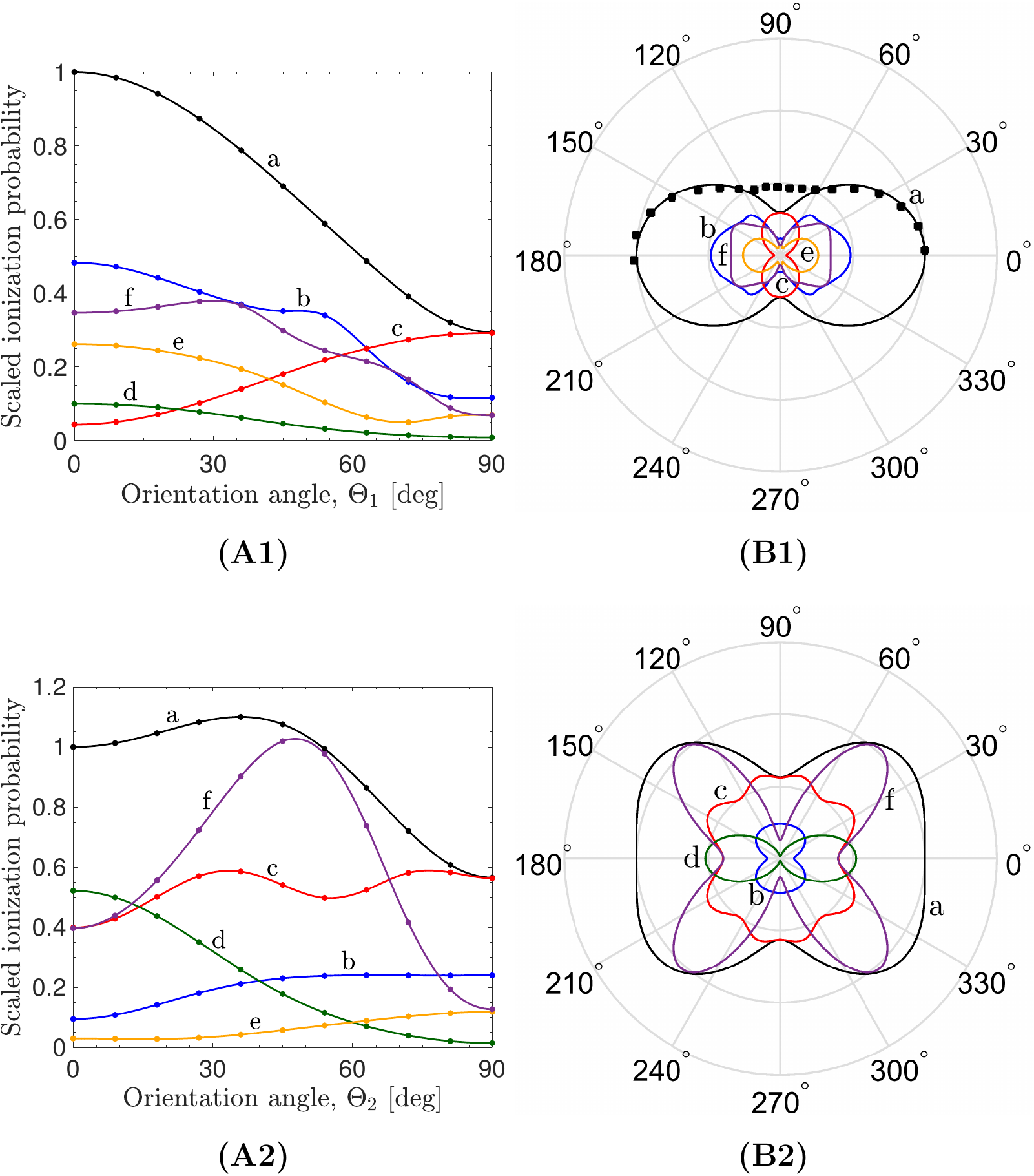}%
}
\caption{\co{Scaled ionization probabilities of benzonitrile molecule, as a function of the orientation angle for linear laser polarizations in the (1) $yz$ and (2) $xz$ planes, with peak intensity of $1 \times 10^{14}$ W/cm$^2$, in Cartesian (panel A) and Polar (panel B) coordinates: (a) total, (b) HOMO, (c) HOMO$-1$, (d) HOMO$-2$, (e) HOMO$-3$ and (f) HOMO$-4$ ionization probabilities. Note that the calculated orbital ionization rates are multiplied by a factor of two and that the ionization rates are scaled to the respective rate at $0^\circ$. The scattered rectangular points correspond to an experimental measurement of the total ionization rate \cite{Hansen2011c}.}} \label{benzonitrile_ionization}
\end{figure}
\indent The breakdown of symmetry and the removal of degeneracy in substituted benzene molecules results in a different angular dependency of the \co{highest-occupied molecular orbital} ionization rates. \co{This is illustrated in Fig. \ref{benzonitrile_ionization} that shows the scaled ionization response of the benzonitrile molecule as a function of the orientation angle in the case of a laser that is linearly polarized in the $yz$ and $xz$ planes, i.e. in the directions $\mathbf{\hat{r}}(\phi=90^\circ,\Theta_1\equiv90^\circ-\theta)$ and $\mathbf{\hat{r}}(\phi=0^\circ,\Theta_2\equiv90^\circ-\theta)$, respectively.} The data is scaled to the total ionization rate at the respective parallel orientation. \co{The} suppression ratio, defined as the ratio between the total ionization yield at the perpendicular and parallel orientations with respect to the \co{molecular plane}, \co{is large compared to that of benzene}. This effect is attributed to the smaller energy difference between the non-degenerate HOMO and HOMO$-1$, i.e. $\simeq 0.28$ eV, and to the larger net permanent dipole of benzonitrile. The ionization response tends to be stronger when the laser polarization is parallel to the \co{molecular plane}, and its dependency on the inner orbitals' contribution is far more significant \co{compared to} benzene. \co{At small $\Theta_1$ angles, the laser polarization is parallel to the nodal $xy$ plane of the HOMO and HOMO-1 $\pi$ molecular orbitals, and the HOMO is dominant due to its lower ionization potential. On the other hand, at small $\Theta_2$ angles the laser polarization is parallel to the nodal $xy$ and the $xz$ planes of the HOMO, and symmetry considerations dictate that the inner orbital response is even stronger than the corresponding ionization of less-bounded orbitals, despite their lower ionization potential. This phenomenon has also been observed in diatomic systems \cite{Telnov2009h}. We note that the ionization of the HOMO$-4$ orbital exceeds the contribution from other orbitals for a wide range of intermediate orientations, e.g. between $\Theta_2 \simeq 15^\circ$ and almost $\Theta_2 \simeq 70^\circ$ for laser polarizations in the $xz$ plane. While the electronic energy of this inner orbital is approximately $2$ eV below that of the HOMO, linear response analysis suggests that it might be strongly excited due to multiphoton resonance with one of the lower-order unoccupied molecular orbitals. Indeed, we found that this orbital ionization is suppressed if the incident laser wavelength is shifted from $800$ nm. Furthermore, it is worth mentioning that the symmetry properties of this orbital are somewhat similar to those of the HOMO, e.g. both are anisotropic along the molecular axis and have a node on the $xz$ plane, cf. Fig. \ref{orbitals}, and, therefore, orbital switching is more likely. Thus approximated models that include fewer high-order occupied orbitals are still able to demonstrate results that compare satisfactorily with experimental measurements, provided that the ionization response accounts for imperfect molecular alignment and orientation (head to tail direction) \cite{Holmegaard2010, Hansen2011, Hansen2011c}. Here, we neglect such orientation effects, and, hence, our fixed-in-space calculations show a dip in the calculated total ionization rate at the perpendicular orientation, i.e. near $\Theta_1 \simeq 90^\circ$ or $\Theta_2 \simeq 90^\circ$, which deviates from the experiment. At this orientation, the relative contribution of the HOMO$-1$ orbital to the ionization response is more significant, which is due to the suppression of HOMO and HOMO$-4$ ionizations at larger projections of the electric field onto the $xz$ nodal surface. Furthermore, the $\sigma$ HOMO$-2$ is also suppressed due to the presence of a nodal plane along the C-CN symmetry axis of the molecule.}\\
\begin{figure}[b]
\centering
\captionsetup[subfigure]{labelformat=empty,position=top,labelfont=bf,textfont=normalfont,singlelinecheck=off,justification=raggedright}
\subfloat[]{\label{first}%
  \includegraphics[width=\linewidth]{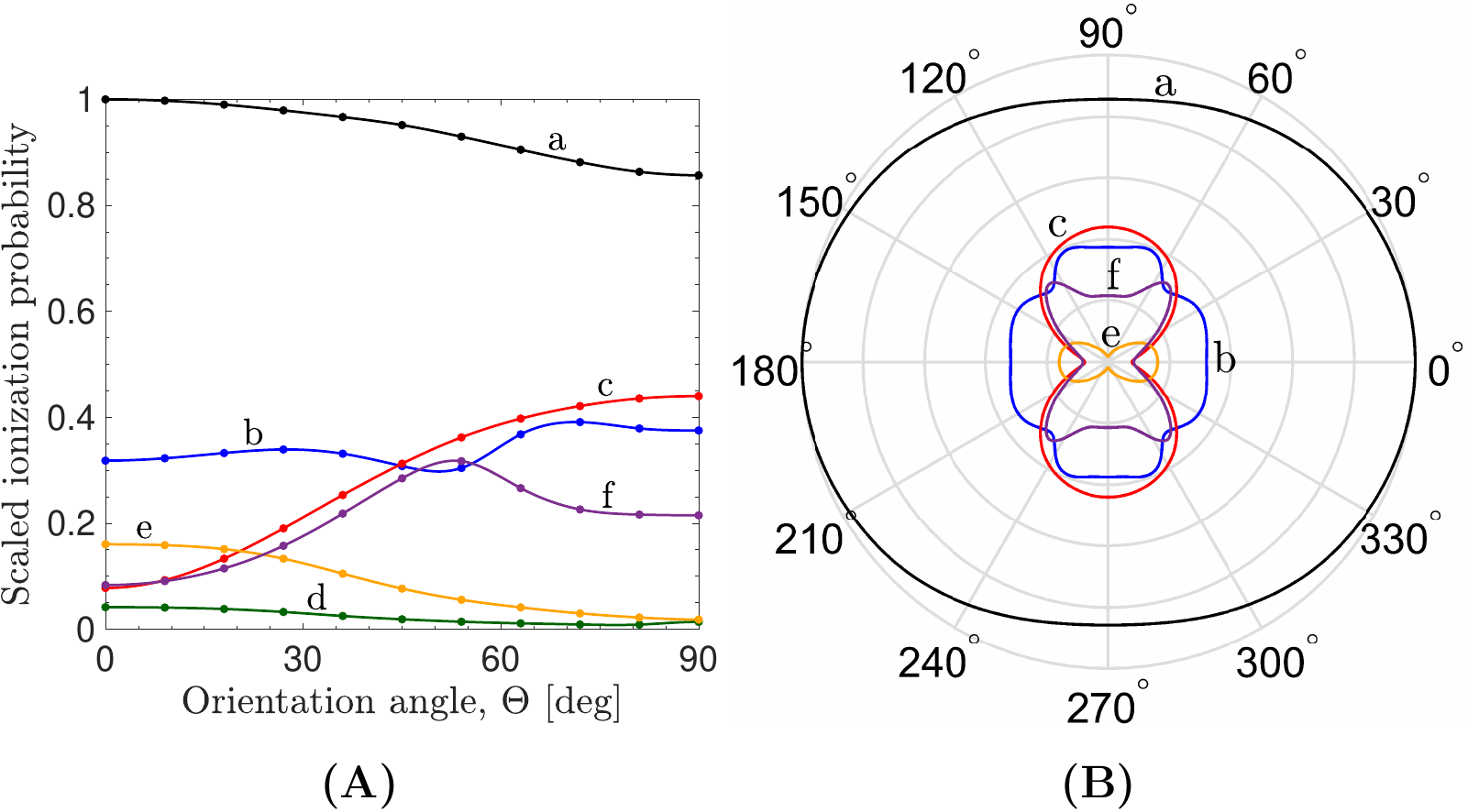}%
}
\caption{Scaled ionization probabilities of fluorobenzene molecule, as a function of \co{the orientation angle $\Theta$, for a linearly polarized laser in the $yz$ plane, with} peak intensity of $3 \times 10^{14}$ W/cm$^2$, in Cartesian (panel A) and Polar (panel B) coordinates: (a) total, (b) HOMO, (c) HOMO$-1$, (d) HOMO$-2$, (e) HOMO$-3$ and (f) HOMO$-4$ ionization probabilities.} \label{fluorobenzene_ionization_high}
\end{figure}
\indent In fluorobenzene, the energy difference between the HOMO and the HOMO$-1$, due to the removal of the degeneracy, is relatively small ($\simeq 0.36$ eV). Thus, the overall ionization yield for a laser with intensity of $1\times 10^{14}$ W/cm$^2$ is only slightly more suppressed \co{compared to the case of} benzene, \co{and shows similar features to Fig. \ref{benzene_ionization}. That is, the suppression of the HOMO near $\Theta \simeq 90^\circ$ is due to the laser polarization being approximately parallel to the nodal surface at $xz$, whereas the HOMO$-1$ is suppressed at $\Theta \simeq 0^\circ$, where the polarization is parallel to the nodes along the molecular and $yz$ planes, in agreement with Ref. \cite{Kjeldsen2005}}. Figure \ref{fluorobenzene_ionization_high} shows the scaled ionization response of fluorobenzene for a \co{laser with a higher} peak intensity of $3 \times 10^{14}$ W/cm$^2$, which is linearly polarized \co{in} the $yz$ plane. Here, the orbital ionization of the HOMO is relatively flat at different \co{molecular orientations}, and it is surpassed by the response of the HOMO$-1$ orbital at $\Theta > 45^\circ$. Furthermore, there is an orbital switching between the response of HOMO and the inner HOMO$-4$ orbital for a narrow range of intermediate laser orientations at around $\Theta \simeq 50^\circ$. \co{We also note that the $\sigma$ orbital angular ionization response is less sensitive to the laser intensity.}\\ 
\indent \co{In principle, the contribution of inner orbitals to the ionization yield becomes more significant on increasing the laser intensity, resulting in a lower suppression ratio and more isotropic total ionization yield.} This is illustrated in Fig. \ref{fluorobenzene_ionization_intensities}, \co{where we} show that the shape of the angle-dependent ionization pattern closely reflects the symmetry of the HOMO orbital \co{for the weakest laser intensity}, with the peak yield obtained here at an orientation angle of $\Theta \simeq 40^\circ$. \co{This result is in agreement with calculations that are based on the strong-field approximation, which only includes incoherent contributions from the HOMO and HOMO$-1$ \cite{Kjeldsen2005}. However, the angular ionization response at higher intensities is flattened, showing lower dependency on the orientation angle. This pattern is in agreement with the calculated ionization response of linear molecules \cite{Son2009g}. We also found that} there is an increase in the relative ionization yield at the perpendicular laser polarization (i.e. for $\Theta \simeq 90^\circ$) due to a more significant contribution from the HOMO$-1$ (cf. Fig. \ref{fluorobenzene_ionization_high}), and the maximum ionization yield shifts to lower orientations.\\
\indent We conclude this section by considering the ionization response of the 1-chloro-2-fluorobenzene molecule. In Fig. \ref{chlorofluorobenzene_ionization} we depict the ionization as a function of the orientation angle $\Theta$, \co{for different linear laser polarizations. Due to the breakdown of the molecular axial symmetry, we consider the orientation angle $\Theta$ to be in the range of $0^\circ - 180^\circ$. We note that, typically, the measured ionization rate under partial 1D alignment conditions corresponds to an average contribution from molecules with different relative orientations between the molecular plane and the laser polarization \cite{Hansen2011c}. The minimum ionization yield is obtained at $\Theta \simeq 90^\circ$, when the laser is linearly polarized in the $yz$ plane, i.e. perpendicular to the molecular plane, cf. Fig. \ref{chlorofluorobenzene_ionization}(a). With the gradual rotation of the laser polarization plane towards the molecular plane, the ionization pattern becomes more flattened and the minimum yield is generally shifted towards smaller orientation angles, cf. Fig. \ref{chlorofluorobenzene_ionization}(b-c). When the plane of laser polarization is parallel to the molecular plane, the ionization yield is minimized at $\Theta \simeq 80^\circ$, cf. Fig. \ref{chlorofluorobenzene_ionization}(d). At this geometry,} the polarization vector is generally directed towards the chlorine atom, in agreement with previous results based on the molecular strong-field approximation \cite{Kjeldsen2005}. \co{The} suppression ratio is somewhat larger in 1-chloro-2-fluorobenzene, \co{compared to} fluorobenzene, despite its having a larger difference between the HOMO and HOMO$-1$ energies, i.e. $\sim 0.49$ eV. This effect is attributed to the contribution of \co{the $\pi$ orbitals} that do not have their nodal planes exactly parallel to the \co{molecular axes} \cite{Kjeldsen2005}.
\begin{figure}
\centering
\captionsetup[subfigure]{labelformat=empty,position=top,labelfont=bf,textfont=normalfont,singlelinecheck=off,justification=raggedright}
\subfloat[]{\label{first}%
  \includegraphics[width=0.950\linewidth]{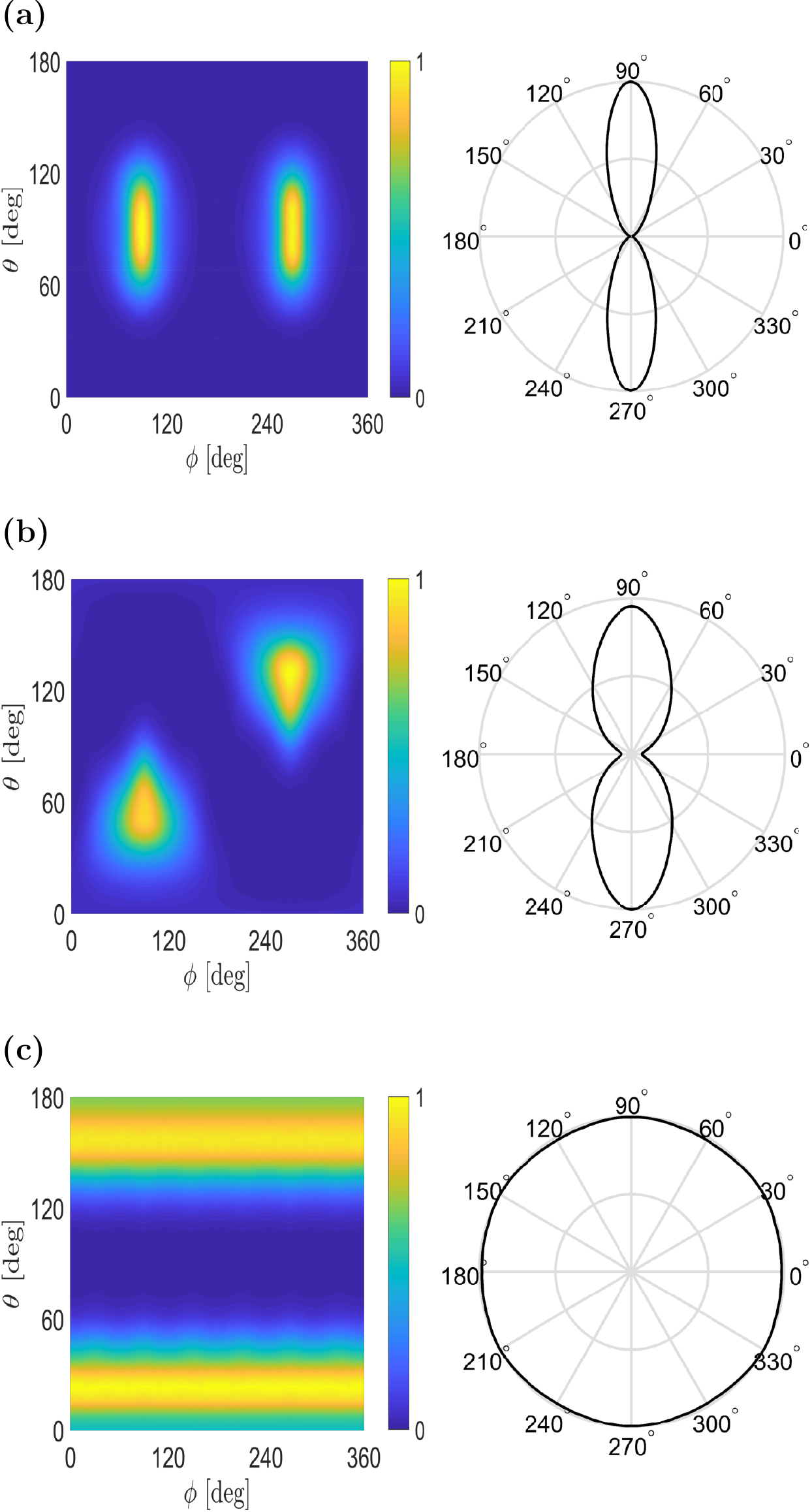}%
}\\
\caption{Photoelectron angular distribution of benzene molecule for different linear laser polarizations in the $yz$ plane: \co{(a) $\Theta=0^\circ$ [$\hat{e}_x=0$, $\hat{e}_y=1$, $\hat{e}_z=0$],} (b) $\Theta=45^\circ$ [0,  $1/\sqrt{2}$,  $1/\sqrt{2}$] and (c) $\Theta=90^\circ$ [0, 0, 1]. The laser wavelength is $800$ nm, with a peak intensity of $1 \times 10^{14}$ W/cm$^2$ and number of optical cycles $N = 10$. The left column shows the normalized PAD distribution on a unit sphere of $r_0 \simeq 10$ a.u., as a function of the spherical coordinates ($\phi$, $\theta$), whereas the right column is the corresponding integrated PAD distribution as a function of the azimuthal angle $\phi$. The color scale corresponds to the level of the normalized ionization flux.}\label{benzene_pad_lin}
\end{figure}
\subsection{Photoelectron Angular Distributions}
\begin{figure}
\centering
\captionsetup[subfigure]{labelformat=empty,position=top,labelfont=bf,textfont=normalfont,singlelinecheck=off,justification=raggedright}
\subfloat[]{\label{first}%
  \includegraphics[width=0.95\linewidth]{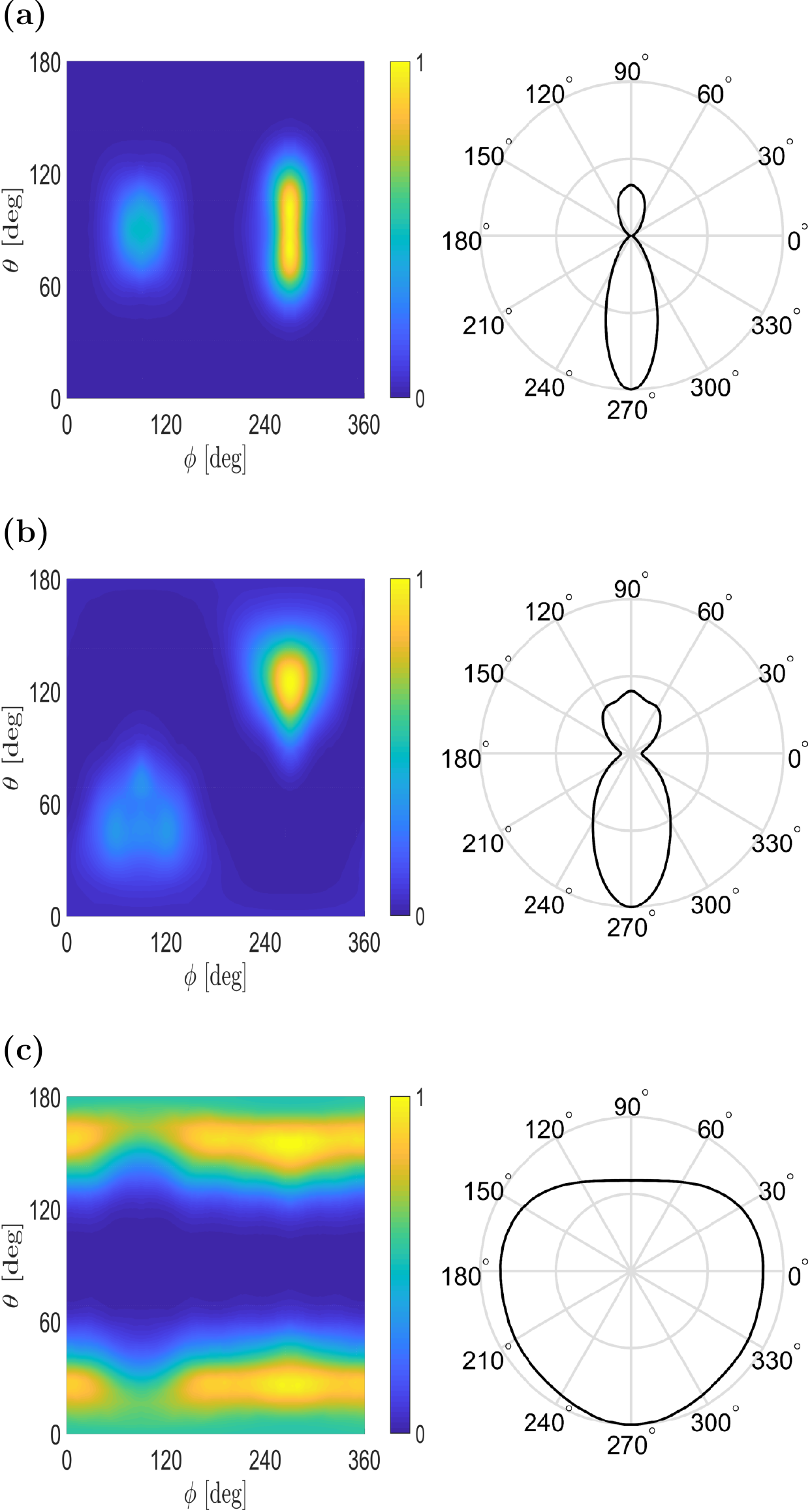}%
}\\
\caption{Same as Fig. \ref{benzene_pad_lin}, but for the fluorobenzene molecule.}\label{fluorobenzene_pad_lin}
\end{figure}

\noindent In this section we consider PADs from substituted benzene molecules, calculated at the end of the laser pulse using Eq. \ref{PAD}, in order to find out how the ionization flux is distributed in space. The PAD plots show the total electron ionization probability on a unit surface sphere (normalized to $r_0$) as a function of the direction defined by the spherical coordinates ($\phi$, $\theta$), cf. Fig. \ref{coordinates}. We also consider the integrated PADs over the angle $\theta$ as a function of the azimuthal angle $\phi$, which preserves the asymmetry features of the full PADs. In Fig. \ref{benzene_pad_lin} we depict PAD plots from a benzene molecule, \co{for different values of the orientation angle $\Theta$ and linear laser polarization in the $yz$ plane, i.e. in the direction $\mathbf{\hat{r}}(\phi=90^\circ, \Theta)$: (a) $\Theta = 0^\circ$, (b) $\Theta = 45^\circ$ and (c) $\Theta = 90^\circ$.} In these plots, the scale of the ionization flux is normalized to unity. The angular ionization flux distribution shows some characteristics that were also found in other molecules \cite{Son2009f}. In particular, \co{we found} that the projected polar PADs for the parallel and intermediate orientations are somewhat similar, characterized by a narrow-center dumbbell shape \co{that} is somewhat thicker in the latter case, whereas the ionization probability is essentially circularly symmetric for the perpendicular orientation. That is, at the parallel orientation \co{($\Theta = 0^\circ$)} the PAD shape is dominated by the contribution of the \co{HOMO(I)} orbital (cf. Fig. \ref{benzene_ionization}) and at the intermediate orientation \co{($\Theta = 45^\circ$)}, the PAD shows an asymmetric peak spot that nearly coincides with the laser polarization. Here, the very small PAD asymmetry is correlated with the initial sign of the electric field. On the other hand, the PAD for the perpendicular orientation \co{($\Theta = 90^\circ$)} is almost free from such phase effects, and it forms a ring shape with a nodal point at its center. \co{This shape} is due to an equal contribution from the two degenerate and orthogonal $\pi$ HOMOs at this orientation.\\
\indent Next, we considered PADs from asymmetric-top molecules, where the existence of a non-negligible permanent net dipole is instrumental to the ionization dynamics \cite{Hansen2011c, Hansen2011}. This is shown in Fig. \ref{fluorobenzene_pad_lin} for fluorobenzene, where we observe a clear enhancement of the ionization flux from the tail of the molecule that is opposite to the fluoride atom, i.e. in the direction of the dipole moment. This enhancement is most significant when the laser is polarized along the direction of the dipole, but it is also apparent in the perpendicular polarization. \co{We note that these PAD asymmetries are not removed by averaging of the carrier envelope phase contributions at different molecular orientations, but, rather, are related to anisotropies in the molecular orbitals distribution \cite{Abu-samha2010b}.}\\ 
\begin{figure*}
\begin{minipage}[c]{0.70\textwidth}
\centering
\captionsetup[subfigure]{labelformat=empty,position=top,labelfont=bf,textfont=normalfont,singlelinecheck=off,justification=raggedright}
\subfloat[]{\label{first}%
  \includegraphics[width=0.95\linewidth]{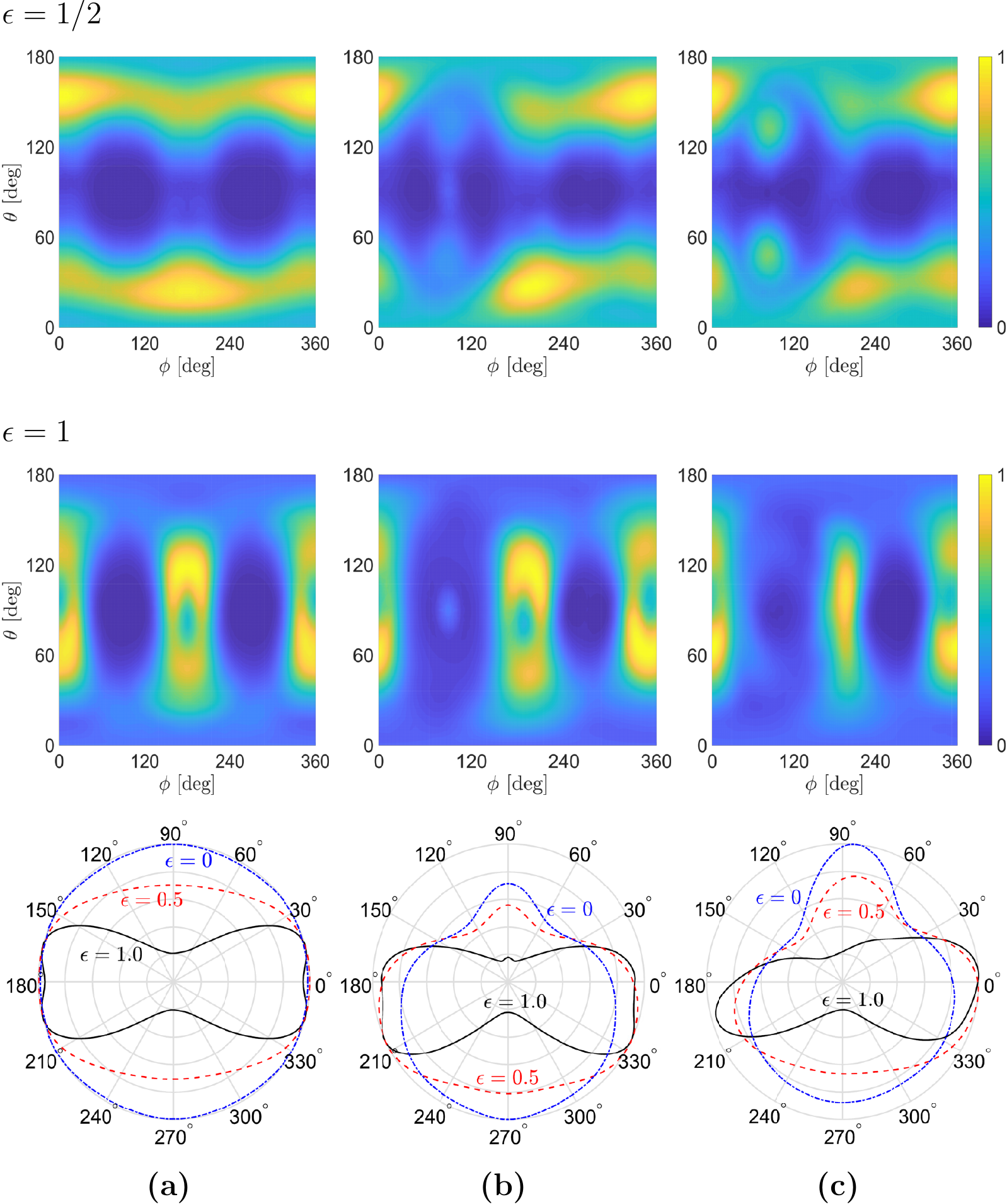}%
}\\
\end{minipage}\hfill
\begin{minipage}[c]{0.30\textwidth}
\caption{Photoelectron angular distributions for three different molecules: (a) benzene , (b) benzonitrile and (c) 1-chloro-2-fluorobenzene. The PADs show the angular ionization distribution as a function of the spherical coordinates ($\phi$, $\theta$), on a unit sphere of $r_0 \simeq 10$ a.u., for a laser pulse polarized on a plane that is parallel to the body-fixed $xz$ plane, with a peak intensity of $3 \times 10^{14}$ W/cm$^2$, N=$10$ optical cycles, and different ellipticity parameters, $\epsilon = 1/2$ (top row) and $\epsilon = 1$ (middle row). The integrated PAD plots in the bottom row represent the normalized ionization flux as a function of the azimuthal angle $\phi$. Here, the case $\epsilon = 0$ corresponds to a linearly polarized laser along the $z$ \co{direction}.}\label{xz_pad}
\end{minipage}
\end{figure*}
\indent Figure \ref{xz_pad} shows PADs for benzene, benzonitrile and 1-chloro-2-fluorobenzene molecules. Here, we studied the effect on the ionization response of laser fields with different degrees of ellipticity, which are polarized on a \co{plane parallel} to the body-fixed \co{$xz$ plane}. In principle, the dynamic of electron ionization and, hence, the overall PAD structure, is largely dependent on the symmetry of the ground-state orbitals that contribute to the ionization process \cite{Martiny2010}. \co{In this setup, the laser polarization (xz plane) is parallel to one of the nodal planes of the HOMO(I) orbital of benzene, whereas it is perpendicular to both nodal planes of the HOMO(II) orbital.} Due to these considerations, \co{the degenerate HOMO(II) orbital contributes most significantly to the ionization process, as} is reflected by the distinguishable lobes that appear in the benzene PAD plots, c.f. Fig. \ref{xz_pad}(a). In particular, the orbital structure is more apparent in the case of the circularly polarized laser ($\epsilon=1$), whereas the ionization in the case of an ellipticly polarized laser ($\epsilon=1/2$) shows contributions from both degenerate HOMOs. Here, the smaller ionization yield at $\theta = 90^\circ$ corresponds to the nodal surface of the HOMOs at the fixed-body \co{molecular plane} \cite{DeGiovannini2012}. \co{Furthermore, the apparent up-down asymmetry in the full PAD, i.e. with respect to this plane, is a consequence of the relative orientation between the molecular axes and the instantaneous polarization of the electric field at its maximum intensity, and, hence, it depends on the initial carrier envelope phase \cite{Holmegaard2010}.} That is, if the helicity of the laser pulse polarization is reversed, i.e. for a right-circularly polarized pulse ($\epsilon=-1$), the up-down asymmetry at a given azimuthal angle is exchanged (not shown). This effect is expected to be small in the calculation of the total ionization rates, and it averages out as the number of optical cycles in the laser pulse is increased \cite{Abu-samha2010b}. \co{The} ionization distribution is symmetric with respect to the center-of mass of the benzene molecule. \co{This pattern} is reflected in the integrated PADs: there is a notable transition from a circular towards a butterfly angular ionization distribution as the polarization of the laser is changed from linear ($\epsilon=0$) to circular ($\epsilon=1$). Similar symmetry arguments also hold for the $\pi$ HOMO and HOMO$-1$ orbitals of benzonitrile. However, the ionization picture is somewhat complicated, \co{compared to} benzene, due to a more significant inner orbital contribution to the ionization yield. \co{That is, we found that the calculated HOMO$-4$ contribution exceeds that of the HOMO and HOMO$-1$ orbitals, in the case of elliptically or circularly polarized laser pulses that have a polarization component along the $x$ axis - in agreement with Fig. \ref{benzonitrile_ionization} (not shown).} In total, there is an ionization enhancement at the tail of the molecule, cf. Fig. \ref{xz_pad}(b). This feature is readily apparent in the asymmetrical shape of the respective integrated PADs. Furthermore, the asymmetrical nature of the HOMO electronic density in benzonitrile is apparent in the PADs, showing a small ionization enhancement towards the head of the molecule, anti-parallel to the direction of the permanent dipole moment (i.e. $\phi \simeq 90^\circ$). This feature is not dependent on the helicity of the laser field in the $xz$ plane and it is more apparent in the case of a linearly polarized laser pulse, where the HOMO contribution to the ionization yield is maximized. In the case of the 1-chloro-2-fluorobenzene molecule, \co{the main contribution to the ionization yield is from the HOMO and HOMO$-1$ orbitals, which are both asymmetrical with respect to the transverse principle body-axes of the molecule,} cf. Fig. \ref{orbitals}. \co{Here,} the corresponding PADs show some enhancement towards the direction of the chlorine atom, cf. Fig. \ref{xz_pad}(c), which is more readily apparent in the case of a linear or elliptic laser polarization, in agreement with the results of Fig. \ref{chlorofluorobenzene_ionization}.\\
\begin{figure*}
\begin{minipage}[c]{0.70\textwidth}
\centering
\captionsetup[subfigure]{labelformat=empty,position=top,labelfont=bf,textfont=normalfont,singlelinecheck=off,justification=raggedright}
\subfloat[]{\label{first}%
  \includegraphics[width=0.95\linewidth]{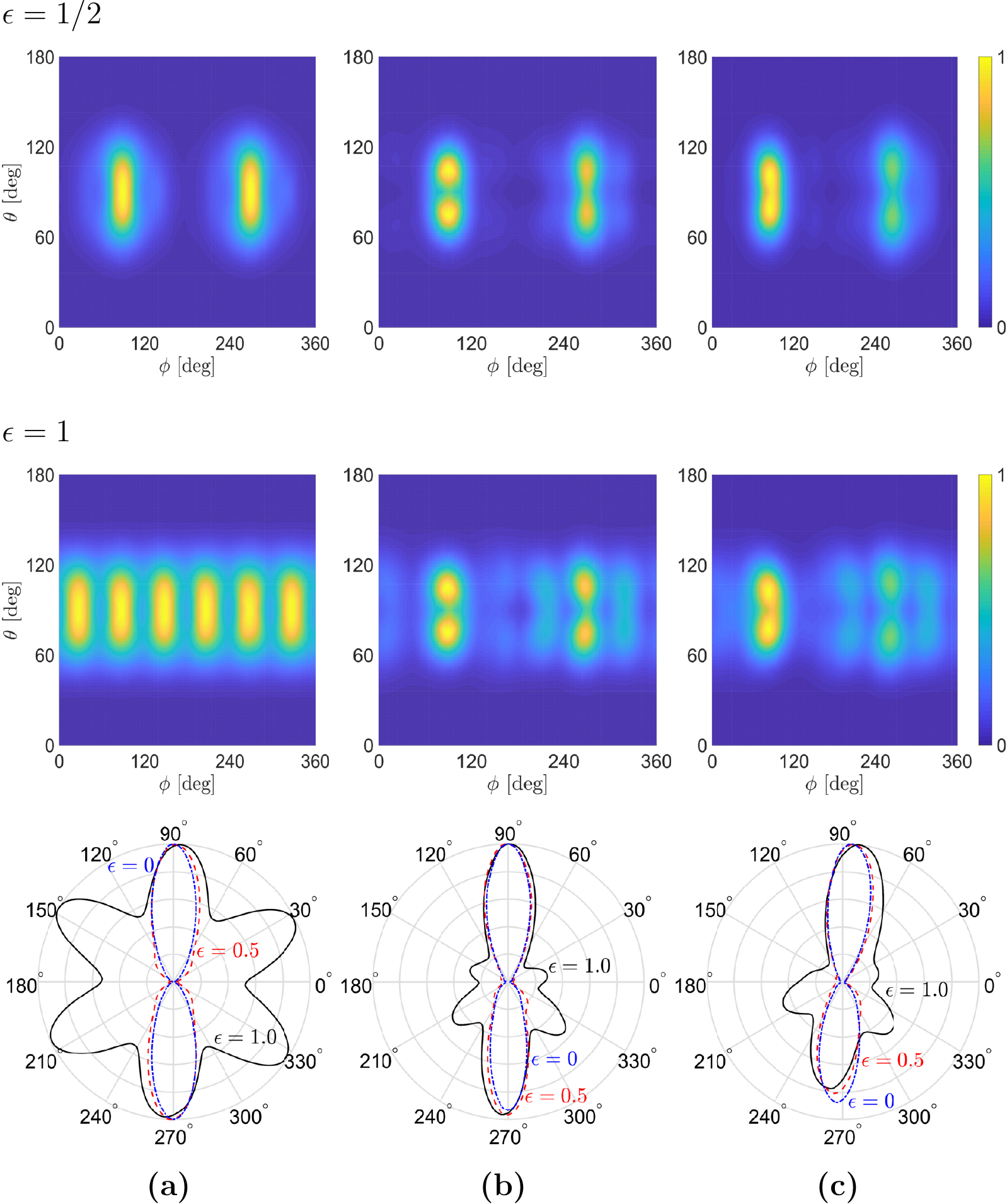}%
}\\
\end{minipage}\hfill
\begin{minipage}[c]{0.30\textwidth}
\caption{Same as Figure. \ref{xz_pad}, but for a laser with a plane of polarization that is parallel to the body-fixed $xy$ plane. Here, the case $\epsilon = 0$ corresponds to a linearly polarized laser along the $y$ axis.}\label{xy_pad}
\end{minipage}
\end{figure*}
\indent Figure \ref{xy_pad} shows normalized PADs for benzene, benzonitrile and 1-chloro-2-fluorobenzene molecules, in the case where the laser pulse is polarized on a plane parallel to the \co{molecular plane}. For this orientation, the common nodal plane of the $\pi$ orbitals coincides with the plane of laser polarization, and thus the complexity of the ionization pattern is somewhat reduced. That is, the field does not have a component perpendicular to this nodal plane, and, therefore, the angular orbital ionization symmetry should be conserved \cite{Dimitrovski2015}. In the case of benzene, the PADs are basically spatially symmetric and the ionization flux is enhanced at the molecule head and tails for linear and elliptic laser polarizations, i.e. $\epsilon = 0$ and $\epsilon = 1/2$, respectively, and at the symmetry points of the benzene ring for a circularly polarized laser pulse ($\epsilon = 1$), cf. Fig. \ref{xy_pad}(a). This dependency is expected from the degenerate HOMOs' structure and their relative contributions to the ionization flux, see Fig. \ref{benzene_ionization}. \co{Here, the electronic density of the molecular orbitals is symmetric with respect to the principle axes, so there is little preference for the ionization from the head ($\phi < 180^\circ$) or tail ($\phi > 180^\circ$) sides of the molecule, except from a minor asymmetry that is related to the carrier envelope phase. On the other hand, the ionization of benzonitrile is dominated by contributions from its anisotropic orbitals, e.g. the HOMO and HOMO$-4$, and the full PADs clearly show an enhancement of the peak ionization flux at the head of the molecule, compared to the yield from its tail.} That is, the ionization enhancement is anti-parallel to the direction pointed to by the permanent \co{molecular dipole}, cf. Fig. \ref{xy_pad}(b). \co{This feature is attributed to the interaction of the laser field with the inhomogeneous electronic charge distribution \cite{Hansen2011, Larsson2016}}. The head and tail ionization asymmetry is present in the integrated PADs, which \co{also} show a sub-structure due to multi-electron contributions as the ellipticity of the laser pulse is increased. \co{This is due to the absence of electronic charge on the $xy$ nodal surface of the $\pi$ orbitals in all benzene derivatives, i.e. at $\theta \simeq 90^\circ$.} \co{We note that the measured electron momentum distribution normal to the molecular plane also shows similar sub-structures, and that such features were found to be strongly dependent on the degree of alignment in the molecular ensemble \cite{Holmegaard2010, Hansen2011}.} The ionization enhancement at the head of the molecule is also apparent in the PADs of 1-chloro-2-fluorobenzene molecule. \co{Here, the fixed-in-space PADs display a small tilt in the orientation of the ionization flux towards the chlorine atom ($\phi < 90^\circ$), which is a consequence of the asymmetrical orbital density redistribution due to the atom substitution, cf. Fig. \ref{orbitals}.} That is, the ionization enhancement in the PADs \co{results from} both the strength and the direction of the permanent dipole moment of the molecule, \co{as it alters} the effective potential of the bound electrons \cite{Abu-samha2010b}. It is also worth mentioning the small left-right asymmetry in the full PADs that is observed on the plane of the laser polarization, i.e. in the direction of the azimuthal angle $\phi$. This feature is much weaker compared to the ionization enhancement due to the dipole-induced effect, and it is mainly dependent on the helicity of the laser pulse \cite{Hansen2011c}.
\section{Conclusions}
\noindent In this work we presented a detailed analysis of the orientation-dependent ionization response of benzene, benzonitrile, fluorobenzene and 1-chloro-2-fluorobenzene molecules using RT-TDDFT. Within this framework, our results take into account the multiple and correlated contributions of inner electronic orbitals to the ionization dynamics in response to an external laser field. Our calculations further extend previous works that were based on the molecular strong-field approximation or limited to the benzene molecule, by taking into account different laser ellipticities, intensity regimes and molecular orientations. We provide additional insights on the differences between the angular ionization distributions of benzene derivatives. The results presented herein can be readily extended to study the orientation dependence and the effect of symmetry breaking on the high-order harmonics generation spectrum of benzene derivatives.\\
\indent We considered the detailed ionization response of the molecules in different orientations with respect to the laser plane of polarization, and for different laser intensities. The ionization response is related to the electronic structure of the molecule and to the symmetries of the molecular orbitals that contribute to the process. The differences in the ionization response are small in the cases of benzene and fluorobenzene molecules, which are characterized by significant contributions from the two \co{highest-occupied molecular orbitals}. The minor changes in the ionization pattern of these molecules is mainly related to the removal of HOMO degeneracy in fluorobenzene. On increasing the laser intensity, the contribution of inner orbitals becomes more significant, resulting in orbital switching at certain molecular orientations, and an alternation of the angular ionization pattern. On the other hand, the benzonitrile molecule is characterized by a larger permanent dipole moment, and the contribution of inner orbitals to the ionization response is enhanced due to resonance conditions at a laser wavelength of 800 nm. \co{This enhancement is particularly strong if the laser is linearly polarized in a plane that is perpendicular to the C-CN symmetry axis of the molecule. We also found that the asymmetrical density distribution in the higher-occupied molecular orbitals of 1-chloro-2-fluorobenzene breaks the rotational symmetry of the ionization pattern, rendering the ionization response dependent on the orientation between the molecular and the laser polarization planes. Our results are in qualitative agreement with previous calculations that were based on molecular ionization theory and the strong-field approximation \cite{Kjeldsen2005}.}\\
\indent We further analysed the PADs of polarizable \co{fluorobenzene} and neutral benzene in the \co{molecular frame}, and showed that the angular resolved ionization pattern reveals important differences that are absent from the full solid angle integrated ionization probabilities. In particular, we found that the ionization response to linear laser polarizations is strongly enhanced towards the tail of the fluorobenzene molecule in the direction of the molecular dipole, whereas it is essentially symmetric in benzene. This effect is most significant when the laser polarization is collinear with the most-polarizable \co{molecular axis}. \co{We also calculated the ionization response with different degrees of laser ellipticity in planes that are either parallel or perpendicular to the molecular plane. Here, we found that the up and down asymmetries in the PADs of asymmetric-top benzonitrile and 1-chloro-2-fluorobenzene molecules depend on the strength and direction of their permanent dipole moment.} We note that the ionization asymmetries are dependent on the electronic structure: we found that in fluorobenzene the PAD is enhanced in alignment, whereas for the other asymmetric-top molecules the enhancement is in anti-alignment to the direction of the permanent \co{molecular dipole}. In the case of the symmetric-top benzene molecule the angular-resolved ionization response is symmetric. \co{Our analysis is in qualitative agreement with experimental observations and calculations that are based on molecular ionization theory, for which integrated PADs were resolved for electrons with momentum components that are perpendicular to the molecular plane \cite{Hansen2011}.} We note, however, that in order to compare more favourably with recent experiments, the fixed-in-space ionization response should be averaged over the temperature-dependent distribution of molecular orientations and should take into account the spatial intensity distribution at the laser beam focus \cite{Hansen2011c}.\\
\section*{Acknowledgments}
AN thanks Prof. Tamar Seideman for useful discussions. AN also wishes to acknowledge support from ISF grant 1722/13. EY wishes to thank Raphael Z. Yahel for making some useful suggestions. \co{We wish to thank an anonymous reviewer for his valuable comments.} This work was also supported by the Cy-Tera project (NEA Y$\Pi$O$\Delta$OMH/$\Sigma$TPATH/0308/31), which is co-funded by the European Regional Development Fund and the Republic of Cyprus through the Research Promotion Foundation.\\

%

\end{document}